\documentclass[11pt]{article}
\pdfoutput=1
\usepackage{fullpage}
\usepackage{xcolor}
\usepackage{cite}
\usepackage{graphicx}
\usepackage{tensor}
\usepackage{amsmath}
\usepackage{amssymb}
\usepackage{subcaption}
\usepackage[utf8x]{inputenc} 
\title{}
\date{}
\renewcommand{\vec}[1]{\mbox{\boldmath$ #1 $}}

\def\beq{\begin{equation}}
\def\eeq{\end{equation}}
\allowdisplaybreaks

\begin{document}
\bibliographystyle{utphys}

\newcommand{\hel}{\eta} 
\renewcommand{\d}{\mathrm{d}}
\newcommand{\dd}{\hat{\mathrm{d}}}
\newcommand{\del}{\hat{\delta}}
\newcommand{\ket}[1]{| #1 \rangle}
\newcommand{\bra}[1]{\langle #1 |}

\newcommand{\be}{\begin{equation}}
\newcommand{\ee}{\end{equation}}
\newcommand\n[1]{\textcolor{red}{(#1)}} 
\newcommand{\diff}{\mathop{}\!\mathrm{d}}
\newcommand{\lb}{\left}
\newcommand{\rb}{\right}
\newcommand{\f}{\frac}
\newcommand{\pd}{\partial}
\newcommand{\tr}{\text{tr}}
\newcommand{\fdiff}{\mathcal{D}}
\newcommand{\im}{\text{im}}
\let\caron\v
\renewcommand{\v}{\mathbf}
\newcommand{\T}{\tensor}
\newcommand{\R}{\mathbb{R}}
\newcommand{\C}{\mathbb{C}}
\newcommand{\Z}{\mathbb{Z}}
\newcommand{\msbar}{\ensuremath{\overline{\text{MS}}}}
\newcommand{\DIS}{\ensuremath{\text{DIS}}}
\newcommand{\abar}{\ensuremath{\bar{\alpha}_S}}
\newcommand{\bb}{\ensuremath{\bar{\beta}_0}}
\newcommand{\rc}{\ensuremath{r_{\text{cut}}}}
\newcommand{\Nd}{\ensuremath{N_{\text{d.o.f.}}}}
\newcommand{\red}[1]{{\color{red} #1}}

\newcommand{\Ad}{\dot{A}}
\newcommand{\Bd}{\dot{B}}
\newcommand{\Cd}{\dot{C}}
\newcommand{\Dd}{\dot{D}}
\newcommand{\Ed}{\dot{E}}
\newcommand{\Fd}{\dot{F}}
\newcommand{\depsilon}{\epsilon}
\newcommand{\dsigma}{\bar{\sigma}}

\newcommand{\bphi}{\phi} 
\newcommand{\bB}{B} 
\newcommand{\bH}{H} 
\newcommand{\bsigma}{\sigma} 
\newcommand{\charge}{\tilde{c}} 
\newcommand{\ampA}{\mathcal{A}} 
\newcommand{\ampM}{\mathcal{M}} 

\titlepage

\vspace*{0.5cm}

\begin{center}
{\bf \Large Non-topological solitons in biadjoint scalar field theory}

\vspace*{1cm} 
\textsc{Kymani Armstrong-Williams\footnote{k.t.k.armstrong-williams@qmul.ac.uk} and
Chris D. White\footnote{christopher.white@qmul.ac.uk}} \\

\vspace*{0.5cm} Centre for Theoretical Physics, School of Physical and
Chemical Sciences, \\ Queen Mary University of London, 327 Mile End
Road, London E1 4NS, UK\\

\end{center}

\vspace*{0.5cm}

\begin{abstract}
  Biadjoint scalar theory has been widely studied, due to its being
  closely related to the double copy correspondence linking gauge,
  gravity and related theories. In this paper, we continue a programme
  of work in elucidating non-linear solutions of this theory, and find
  a family of new solutions that are richer and more complex than
  previous cases. Using an ansatz that can be embedded in any choice
  of non-abelian colour groups, we demonstrate the existence of
  non-topological solitons, whose existence is protected by carrying a
  U(1) charge associated with certain rotations in colour space. The
  solutions are time-dependent, and closely related to the well-known
  Q-ball solutions in other scalar field theories. We also show
  explicitly that our solution set contains those that are stable
  under small perturbations within a consistent truncation of the
  theory, and have finite energy in addition to being localised.
\end{abstract}

\vspace*{0.5cm}

\section{Introduction}
\label{sec:intro}

In recent years, a relationship known as the {\it double copy} has
been studied, that acts as a precise map between scattering
amplitudes~\cite{Kawai:1985xq,Bern:2008qj,Bern:2010ue,Bern:2010yg} and
classical
solutions~\cite{Monteiro:2014cda,Luna:2015paa,Ridgway:2015fdl,Bahjat-Abbas:2017htu,Carrillo-Gonzalez:2017iyj,CarrilloGonzalez:2019gof,Bah:2019sda,Alkac:2021seh,Alkac:2022tvc,Luna:2018dpt,Sabharwal:2019ngs,Alawadhi:2020jrv,Godazgar:2020zbv,White:2020sfn,Chacon:2020fmr,Chacon:2021wbr,Chacon:2021hfe,Chacon:2021lox,Dempsey:2022sls,Easson:2022zoh,Chawla:2022ogv,Han:2022mze,Armstrong-Williams:2022apo,Han:2022ubu,Elor:2020nqe,Farnsworth:2021wvs,Anastasiou:2014qba,LopesCardoso:2018xes,Anastasiou:2018rdx,Luna:2020adi,Borsten:2020xbt,Borsten:2020zgj,Goldberger:2017frp,Goldberger:2017vcg,Goldberger:2017ogt,Goldberger:2019xef,Goldberger:2016iau,Prabhu:2020avf,Luna:2016hge,Luna:2017dtq,Cheung:2016prv,Cheung:2021zvb,Cheung:2022vnd,Cheung:2022mix,Chawla:2024mse,Keeler:2024bdt,Chawla:2023bsu,Easson:2020esh,Armstrong-Williams:2024bog,Armstrong-Williams:2023ssz,Farnsworth:2023mff,Morieri:2026gdo,Aoude:2025jvt,Alkac:2026xbp,Didenko:2026nip,Easson:2026wex,Misuna:2026has,Rodriguez:2026zxm,Easson:2026xod,Garcia-Compean:2026cnq,Carrasco:2025bgu,Ilderton:2025aql}
in gauge, gravity and related theories (see
refs.~\cite{Borsten:2020bgv,Bern:2019prr,Adamo:2022dcm,Bern:2022wqg,White:2021gvv,White:2024pve}
for pedagogical reviews). Potential non-perturbative aspects have been
explored in
refs.~\cite{Monteiro:2011pc,Borsten:2021hua,Alawadhi:2019urr,Banerjee:2019saj,Huang:2019cja,Berman:2018hwd,Alfonsi:2020lub,Alawadhi:2021uie,White:2016jzc,DeSmet:2017rve,Bahjat-Abbas:2018vgo,Cheung:2022mix,Moynihan:2021rwh,Borsten:2022vtg,Holton:2026eha,Carrasco:2026ijt,Albertini:2025ogf,Holton:2025hny},
although the ultimate scope and remit of the double copy remain
somewhat mysterious. In all cases where the correspondence has been
studied, a novel field theory known as {\it biadjoint scalar field
  theory} plays a crucial role. The central object in this theory is a
scalar field with two adjoint (colour) indices, such that associated
particles carry two different types of colour charge. When turning
gauge theory amplitudes into gravity ones, for example, one leaves
untouched the denominators, which in turn can be interpreted as scalar
propagators. For classical solutions, the biadjoint field again shows
up as that part of a solution that is left untouched when constructing
gravity solutions from gauge theory building blocks. However, in all
previous cases where exact or perturbative solutions have been copied,
the relevant biadjoint fields satisfy the {\it linearised} equations
of motion. This may in turn explain why a true non-perturbative
incarnation of the double copy has yet to be found.

The above remarks suggest that one route to extending the double copy
may be to find non-linear solutions of biadjoint scalar field theory,
with the hope that they may act as building blocks for constructing
non-linear gauge and / or gravity solutions. To this end,
refs.~\cite{White:2016jzc,DeSmet:2017rve,Bahjat-Abbas:2018vgo,Armstrong-Williams:2025spu,Armstrong-Williams:2022apo}
have found an expanding catalogue of non-perturbative solutions, of
increasing complexity. In refs.~\cite{White:2016jzc,DeSmet:2017rve},
spherically symmetric monopole-like solutions were presented which,
although they can be dressed by a non-trivial profile function, remain
singular. Reference~\cite{Bahjat-Abbas:2018vgo} performed a similar
analysis for axisymmetric solutions, and
ref.~\cite{Armstrong-Williams:2022apo} attempted to find non-linear
solutions in Euclidean signature, with a negative result. More
recently, ref.~\cite{Armstrong-Williams:2025spu} found time-dependent
solutions for the first time, in the form of travelling waves with a
non-trivial non-linear profile. These also included bounded solutions,
suggesting that the non-perturbative structure of biadjoint theory is
richer than previously thought. The authors also included quartic
extensions of the standard cubic biadjoint theory, which have been
motivated in previous double-copy
studies~\cite{Banerjee:2018tun,Kalyanapuram:2019nnf,Aneesh:2019cvt,Srivastava:2020dly}.

The aim of this paper is to continue the above programme of work, and
to present an interesting new family of solutions for (in general
quartic) biadjoint scalar field theory. These take the form of
spherically symmetric time-dependent solutions, which have a number of
novel features compared to previous works. In particular, they are
localised oscillatory objects, with finite profiles and hence finite
energy. Furthermore, the two colour groups can be chosen
independently, in contrast with the more restricted choices that are
inherent in
refs.~\cite{Banerjee:2018tun,Kalyanapuram:2019nnf,Aneesh:2019cvt,Srivastava:2020dly}. We
will further see that the new solutions carry a conserved charge $Q$,
that acts to protect their existence. This makes them a form of
non-topological
soliton~\cite{Lee:1974ma,Friedberg:1976me,Friedberg:1976az,Friedberg:1976ay,Friedberg:1976eg,Friedberg:1977xf}
(see e.g. ref.~\cite{Lee:1991ax} for a review), and indeed they appear
to be closely related to the well-known {\it Q-balls} of
ref.~\cite{Coleman:1985ki}. As such, we will be able to determine the
existence of a discrete spectrum of higher states, and also explicitly
show that the lowest energy state is stable within a consistent
truncation of the theory. Whilst the matching of non-linear solutions
between biadjoint scalar field theory and more physical theories
remains an open problem, our results provide the most compelling
evidence yet that there is a diverse zoo of localised objects living
inside biadjoint theories, which may well find applications elsewhere.

The structure of our paper is as follows. In section~\ref{sec:ansatz},
we review the field equations for biadjoint scalar theory, and explain
the ansatz we have used to obtain our new solutions. In
section~\ref{sec:sols}, we demonstrate the existence of
non-topological solitons, and describe the conditions in which they
appear. In section~\ref{sec:stability}, we analyse the stability of
our solutions, noting in particular the role of the soliton
charge. Finally, we discuss our results and conclude in
section~\ref{sec:discuss}.

\section{Biadjoint scalar field theory}
\label{sec:ansatz}

In this section, we review salient details regarding biadjoint scalar
field theory, and introduce a particular solution ansatz that
underpins the results of this paper. Our starting point is to consider
the field
\begin{equation}
  {\bf \Phi}=\Phi^{aa'}{\bf T}^a\otimes \tilde{\bf T}^{a'},
  \label{Phidef}
\end{equation}
valued in two Lie algebras with generators $\{{\bf T}^a\}$ and $\{{\bf
  T}^{a'}\}$ associated with groups $G$ and $\tilde{G}$, satisfying
the commutation relations
\begin{equation}
  [{\bf T}^a,{\bf T}^b]=if^{abc}{\bf T}^c,\quad
  [\tilde{\bf T}^{a'},\tilde{\bf T}^{b'}]
  =i\tilde{f}^{a'b'c'}\tilde{\bf T}^{c'}.
  \label{comms}
\end{equation}
The field equation for the components $\Phi^{aa'}$ can then be written
as
\begin{equation}
  \Box \Phi^{aa'}+y f^{abc}\tilde{f}^{a'b'c'}\Phi^{bb'}\Phi^{cc'}
  -gf^{ebc}f^{eda}\tilde{f}^{e'b'c'}\tilde{f}^{e'd'a'}\Phi^{bb'}
  \Phi^{cc'}\Phi^{dd'}=0.
  \label{EOM}
\end{equation}
For convenience in what follows, we have adopted different conventions
to e.g. ref.~\cite{Armstrong-Williams:2025spu}, also choosing to write
the quartic term in terms of a single colour structure, rather than a
symmetrised combination of colour structures (n.b. the latter may be
shown to be equivalent to the former). Furthermore, we will work in
mostly plus signature, with d'Alambertian
$\Box=-\partial_t^2+\nabla^2$ in terms of the spatial Laplacian
$\nabla^2$.

In ref.~\cite{White:2016jzc}, it was pointed out that solutions of the
form
\begin{equation}
  \Phi^{aa'}(x)= c^a\tilde{c}^{a'}\phi(x),
  \label{Phiansatz1}
\end{equation}
where $c^a$ and $\tilde{c}^{a'}$ are constant colour vectors, are such
as to linearise the field equations. Non-trivial solutions were then
obtained by replacing the outer product of two colour vectors with a
higher-rank matrix e.g.
\begin{equation}
  \Phi^{aa'}(x)= \delta^{aa'}\phi(x).
  \label{Phiansatz2}
\end{equation}
However, such ans\"{a}tze have the effect of constraining the choice
of gauge groups, where the example of eq.~(\ref{Phiansatz2}) is such
that the dimensions of $G$ and $\tilde{G}$ must match. Here, we take a
different approach, motivated by previous work in Yang-Mills
theory~\cite{Smilga:2001ck}. That is, we assume the ansatz
\begin{equation}
  \Phi^{aa'}=\sum_{i=1}^M c_i^a \tilde{c}_i^{a'}\phi_i,
  \label{Phiansatz3}
\end{equation}
involving a sum over $M$ outer products of colour vectors, each of
which is weighted by a potentially different scalar field
$\phi_i(x)$. This ansatz is more general than any previously studied
solution, and indeed includes eq.~(\ref{Phiansatz2}) as a special
case. To see this, one may restrict the dimensions of the two groups
$G$ and $\tilde{G}$ to be the same value $N$, and set $M=N$. Upon
choosing $\phi_i\equiv \phi$ to be a common scalar field, and the
colour vectors to be given by $c_i^{a}=\delta_i^a$ and similarly for
$\tilde{c}_i^{a'}$, eq.~(\ref{Phiansatz3}) reduces to
\begin{equation}
  \Phi^{aa'}\rightarrow \phi\sum_{i=1}^N \delta_i^a\delta_i^{a'}=
  \delta^{aa'}\phi.
  \label{Phiansatz3lim}
\end{equation}
Returning to the general case, the role of the colour vectors in
eq.~(\ref{Phiansatz3}) is to pick out particular combinations of
colour generators in each group. Indeed, defining the
Lie-algebra-valued matrices
\begin{equation}
  {\bf c}_i=c_i^a{\bf T}^a,\quad \tilde{\bf c}_i=\tilde{c}_i^{a'}{\bf T}^{a'},
  \label{cmat}
\end{equation}
the full biadjoint field of eq.~(\ref{Phidef}) becomes, with the
ansatz of eq.~(\ref{Phiansatz3}),
\begin{equation}
  {\bf \Phi}(x)=\sum_{i=1}^M {\bf c}_i\otimes \tilde{\bf c}_i\,\phi_i(x).
\label{Phiansatz4}
\end{equation}
One may then write the field equation directly in terms of these
quantities, by contracting eq.~(\ref{EOM}) with ${\bf T}^a\otimes{\bf
  T}^{a'}$, and using eq.~(\ref{comms}):
\begin{equation}
  \sum_{i}{\bf c}_i\otimes \tilde{\bf c}_i
  \Box \phi_i-2y\sum_{i<j} [{\bf c}_i,{\bf c}_j]\otimes
    [\tilde{\bf c}_i,\tilde{\bf c}_j]\phi_i\phi_j
    -2g\sum_{i<j}\sum_k \left[[{\bf c}_i,{\bf c}_j],{\bf c}_k\right]
    \otimes
    \left[[\tilde{\bf c}_i,\tilde{\bf c}_j],\tilde{\bf c}_k\right]
    \phi_i\phi_j\phi_k=0.
\label{EOM2}
\end{equation}
The various (nested) commutators can be evaluated upon choosing a
suitable form for $\{{\bf c}_i,\tilde{\bf c}_i\}$, for which different
choices constitute different solutions\footnote{Whether or not
different solutions are physically distinct depends upon whether their
respective colour vectors are related by (global) gauge
transformations.}. A particularly well-motivated choice, that can be
made in any non-abelian semisimple Lie algebra, is
\begin{equation}
  ({\bf c}_1,{\bf c}_2,{\bf c}_3)=({\bf E}_+,{\bf E}_-,{\bf H}),\quad
  (\tilde{\bf c}_1,\tilde{\bf c}_2,\tilde{\bf c}_3)
  =(\tilde{\bf E}_+,\tilde{\bf E}_-,\tilde{\bf H}).
  \label{c123}  
\end{equation}
Here ${\bf H}$ is a member of the Cartan subalgebra of the Lie algebra
of $G$. The pair $[{\bf E}_+,{\bf E}_-]$ is associated with a
so-called {\it root} of the Lie
algebra~\cite{Georgi:1999wka,Zee:2016fuk}, such that the following
commutation relations are obeyed\footnote{We note that different
conventions are sometimes used, such that the factor of 2 in
eq.~(\ref{comms2}) appears in the first commutator rather than the
second. This amounts to the refinition $H\rightarrow 2H$.}:
\begin{equation}
  [{\bf E}_+,{\bf E}_-]={\bf H},\quad [{\bf H},{\bf E}_\pm]=\pm 2{\bf E}_\pm,
  \label{comms2}
\end{equation}
and similarly for the tilded quantities. Substituting eq.(\ref{c123})
into eq.~(\ref{EOM2}) and repeatedly using eq.~(\ref{comms2}) yields
\begin{align}
  &{\bf E}_+\otimes\tilde{\bf E}_+\Big[\Box \phi_1-8y\phi_1\phi_3
    -32g\phi_1\phi_3^2-8g\phi_1^2\phi_2
    \Big]
  +{\bf E}_-\otimes \tilde{\bf E}_-\Big[\Box \phi_2-8y\phi_2\phi_3
    -32g\phi_2\phi_3^2-8g\phi_1\phi_2^2
    \Big]\notag\\
  &\quad+{\bf H}\otimes\tilde{\bf H}\Big[\Box\phi_3-2y\phi_1\phi_2
    -16g\phi_1\phi_2\phi_3
    \Big]=0.
  \label{EOM3}
\end{align}
By the linear independence of each of the colour generator
combinations, we may now set the contents of each square bracket to
zero, which gives the following set of non-linear coupled partial
differential equations:
\begin{align}
  \Box\phi_1-8y\phi_1\phi_3-32g\phi_1\phi_3^2-8g\phi_1^2\phi_2&=0;\notag\\
  \Box\phi_2-8y\phi_2\phi_3-32g\phi_2\phi_3^2-8g\phi_1\phi_2^2&=0;\notag\\
  \Box\phi_3-2y\phi_1\phi_2-16g\phi_1\phi_2\phi_3&=0.
  \label{phi123}
\end{align}
These possess the global symmetry
\begin{equation}
  \phi_1\rightarrow e^{i\alpha}\phi_1,\quad \phi_2\rightarrow e^{-i\alpha}
  \phi_2,\quad \phi_3\rightarrow \phi_3,
  \label{phitrans}
\end{equation}
corresponding to complex rotations of the fields
$(\phi_1,\phi_2)$. The origin of this symmetry is that the form of the
ansatz of eqs.~(\ref{Phiansatz4}, \ref{c123}) is left invariant after
performing complex rotations about the ${\bf H}$ direction in colour
space, which in turn rotates the generators ${\bf E}_+$ and ${\bf
  E}_-$ into each other.

\section{Existence of non-topological solitons}
\label{sec:sols}

Having derived the set of coupled equations in eq.~(\ref{phi123}), let
us now set about solving them. To this end, we make the further ansatz
\begin{equation}
  \phi_i(x)=R_i(r)e^{i\omega_i t},
  \label{phiR}
\end{equation}
such that each scalar field appearing in eq.~(\ref{Phiansatz4}) has an
oscillatory behaviour, with an amplitude depending on the radial
coordinate $r$. Substituting this into eq.~(\ref{phi123}) and
requiring that the time dependence matches on both sides of each
equation yields the constraints
\begin{equation}
  \omega_1=-\omega_2\equiv \omega,\quad \omega_3=0.
  \label{omegavals}
\end{equation}
Given the symmetric role of $\phi_1$ and $\phi_2$, we may furthermore
focus on the symmetric choice $R_1(r)=R_2(r)$, in which case
eqs.~(\ref{phi123}) reduce to
\begin{align}
  (\omega^2+\nabla^2)R-8yRR_3-32gRR_3^2-8gR^3&=0;\notag\\
  \nabla^2 R_3-2yR^2-16gR^2 R_3=0.
  \label{RR3}
\end{align}
To further interpret these equations, it is convenient to shift $R_3$
by writing
\begin{equation}
  R_3(r)=S(r)+c,
  \label{R3shift}
\end{equation}
where
\begin{equation}
  c=\frac{-y+\beta}{8g},\quad \beta=\sqrt{y^2+2g\omega^2}
  \label{cdef}
\end{equation}
is chosen to explicity remove the $\omega^2$ term in the first
equation of eq.~(\ref{RR3}). The latter equations become
\begin{align}
  \nabla^2 R-8\beta S R-32gRS^2-8gR^3&=0;\notag\\
  \nabla^2S-16gR^2S-2\beta R^2&=0.
  \label{RS}
\end{align}
We may now look for localised solutions by requiring that $R$ falls
off sufficiently quickly as $r\rightarrow \infty$, where we will
further assume that $S(r)$ tends to a constant $S_\infty$. In the
large $r$ regime where higher-order terms in $R$ can be neglected,
eq.~(\ref{RS}) reduces to
\begin{equation}
  (\nabla^2 -\mu^2)R(r)=0,\quad \nabla^2 S(r)=0,
  \label{RS2}
\end{equation}
where we have defined the effective squared mass parameter
\begin{equation}
  \mu^2=8\beta S_\infty +32g S_\infty^2.
  \label{mu2def}
\end{equation}
Provided this parameter is positive, there will indeed be localised
solutions, which fall off in a Yukawa-type fashion at large
$r$. Together with the fact that $S$ becomes harmonic in this limit,
the asymptotic behaviour of both functions is 
\begin{equation}
  R(r)\sim \frac{e^{-\mu r}}{r},\quad S(r)\sim S_{\infty}-\frac{K}{r},
\label{RSinf}
\end{equation}  
for some constant $K$. Below, we will explicitly verify that such
solutions exist, which are regular at the origin. Before doing so, it
is worth commenting on why such objects are stable against decay to
the trivial vacuum solution $\phi_1=\phi_2=\phi_3=0$, and the answer
is that they carry a conserved charge, associated with the U(1)
symmetry for rotations about the ${\bf H}$ direction, noted at the end
of the previous section. Our choice of $R_1=R_2$ means that one may
write
\begin{equation}
  \phi\equiv \phi_1=\phi_2^*=R(r)e^{i\omega t},
  \label{phidef}
\end{equation}
such that eq.~(\ref{phi123}) becomes
\begin{align}
  \Box \phi-8y\phi\phi_3-32g\phi\phi_3^2-8g|\phi|^2\phi&=0;\notag\\
  \Box \phi_3-2y|\phi|^2-16g|\phi|^2\phi_3&=0,
  \label{EOMphi}
\end{align}
where the second equation in eq.~(\ref{phi123}) can be obtained as the
complex conjugate of the first equation of eq.~(\ref{EOMphi}). These
equations can be obtained from the reduced Lagrangian\footnote{The
minus sign in the kinetic terms in eq.~(\ref{Ldef}) arises from our
use of mostly plus signature. Furthermore, the non-canonical
normalisation of the $\phi_3$ field arises from our choice of
conventions in defining the generator combinations in
eq.~(\ref{c123}).}
\begin{equation}
  {\cal L}=-(\partial^\mu\phi^*)(\partial_\mu\phi)
  -2(\partial^\mu\phi_3)(\partial_\mu\phi_3)-8y|\phi|^2\phi_3
  -32g|\phi|^2\phi_3^2-4g|\phi|^4.
  \label{Ldef}
\end{equation}
The U(1) symmetry is here directly visible as $\phi\rightarrow
e^{i\alpha}\phi$, and the associated conserved Noether current is
\begin{equation}
  j^\mu=i\left(\phi^*\partial^\mu\phi-\phi\partial^\mu\phi^*\right).
  \label{jmudef}
\end{equation}
The associated charge density, integrated over all space, is then
\begin{equation}
  Q=i\int d^3\vec{x}\left[\phi\dot{\phi}^*-\phi^*\dot{\phi}\right]
  =8\pi\omega \int_0^\infty dr r^2 R^2(r).
  \label{Qdef}
\end{equation}
Conservation of this charge forbids solutions from transitioning to
states with different (and in particular lower) $Q$. Thus, the
existence of such solutions is protected, in a way that is different
to solitons that are protected by having a non-vacuum topology at
infinity. It is for this reason that solutions of the type discussed
here are referred to as {\it non-topological solitons}. Whilst
protected, they can still be unstable, as there may be an ability to
transition to a state with lower energy, but with the same charge. We
return to the issue of stability in section~\ref{sec:stability}, but
it is worth commenting here that one can understand the ansatz of
eq.~(\ref{phidef}) as arising from the condition of minimising the
energy for a given fixed charge $Q$. Given the Lagrangian of
eq.~(\ref{Ldef}), the energy is
\begin{equation}
  E=\int d^3\vec{x} \Big[
    |\dot{\phi}|^2+2\dot{\phi}_3^2+|\nabla\phi|^2
    +2|\nabla\phi_3|^2
    +V(\phi,\phi_3)
    \Big],
  \label{Edef}
\end{equation}
where we have introduced the potential
\begin{equation}
  V(\phi,\phi_3)=8y|\phi|^2\phi_3+32g|\phi|^2\phi_3^2+4g|\phi|^4.
  \label{Vdef}
\end{equation}
To minimise the energy for a given $Q$, we can introduce a Lagrange
multipler $\omega$, and consider the functional
\begin{align}
  E_\omega&=E+\omega\left(Q-i\int d^3\vec{x}(\phi\dot{\phi}^*-\phi^*\dot{\phi})
  \right)\notag\\
  &=\int d^3\vec{x}\Big[
    |\dot{\phi}-i\omega \phi|^2+2\dot{\phi}_3^2+|\nabla \phi|^2
    +2|\nabla\phi_3|^2+V(\phi,\phi_3)-\omega^2|\phi|^2
    \Big]+\omega Q,
  \label{Eomega}
\end{align}
where we have used eq.~(\ref{Edef}) and rearranged in the second
line. The only time dependence appears in the first two terms of the
second line, which are positive definite. Thus, minimising the
functional implies
\begin{equation}
  \dot{\phi}=i\omega \phi,\quad \dot{\phi_3}=0,
  \label{phiset}
\end{equation}
which indeed (for spherically symmetric solutions) leads to the ansatz
of eq.~(\ref{phidef}) and the constraints of eq.~(\ref{omegavals}). We
therefore see that the Lagrange multiplier takes on a physical
interpretation, as the oscillation frequency of the solutions. Such
arguments are well-known in the context of other scalar theories,
where the analogous solutions are known as Q-balls after
ref.~\cite{Coleman:1985ki}. The solutions here can perhaps be viewed
as novel generalisations of these: there are two fields rather than
one. Also, standard Q-balls exist in scalar theories with a mass term,
where a certain condition must be obeyed between the perturbative mass
$m$ and the frequency $\omega$. In the present case, a similar
condition arises, even though the perturbative mass is zero. We saw
above that an effective mass parameter $\mu$ arises, such that
localised solutions exist for $\mu^2>0$. From eq.~(\ref{mu2def}), the
origin of this effect is that the field $S(r)$ (defined in
eqs.~(\ref{R3shift}, \ref{cdef}) in terms of $\phi_3$) approaches a
constant at large $r$. This constant is not a free choice, but is
determined dynamically by the consistent solution of the equations of
motion for the coupled fields $\phi$ and $\phi_3$. Thus, it is the
non-trivial interplay between the different components of the
biadjoint field that creates the possibility for localised
non-topological solitons to exist.

Having discussed the role of the charge, let us now turn to finding
explicit forms for the radial functions $R(r)$ and $S(r)$. The
solution of eq.~(\ref{RS}) is not analytically tractable, but one may
instead solve the equations of motion numerically using a standard
shooting method. One starts by imposing boundary values at $r=0$:
\begin{equation}
  R(0)=a,\quad S(0)=b.
  \label{R0S0}
\end{equation}
Regularity at the origin further implies
\begin{equation}
  R'(0)=0,\quad S'(0)=0.
  \label{R0S02}
\end{equation}
Next, one can numerically integrate the differential equations out to
some large radius $r$, where the aim is to find the precise value of
$b$ such that the exponentially growing mode $R\sim e^{\mu r}/r$ is
absent. Given that this mode may occur with positive or negative
coefficient, choosing the wrong value of $b$ will lead to the
numerical solution either blowing up to positive, or negative,
infinity. One may thus start with a range of $b$ values, and perform
bisection until the critical value $b=b_*$ is reached corresponding to
a localised solution. For a given set of parameters $(y,g,\omega,a)$,
a discrete spectrum of $b_*$ values is obtained, corresponding to
different solutions with the same $\omega$. Given that these are
stationary points of the function $E_\omega$ which depends on both the
charge and energy, fixing $\omega$ does not correspond to fixing
either the charge $Q$ or energy $E$. Instead, the spectrum of states
for each $\omega$ consists of states with increasing numbers of nodes
in the radial function $R(r)$, and increasing energy / charge.

\begin{figure}
  \begin{center}
    \scalebox{0.6}{\includegraphics{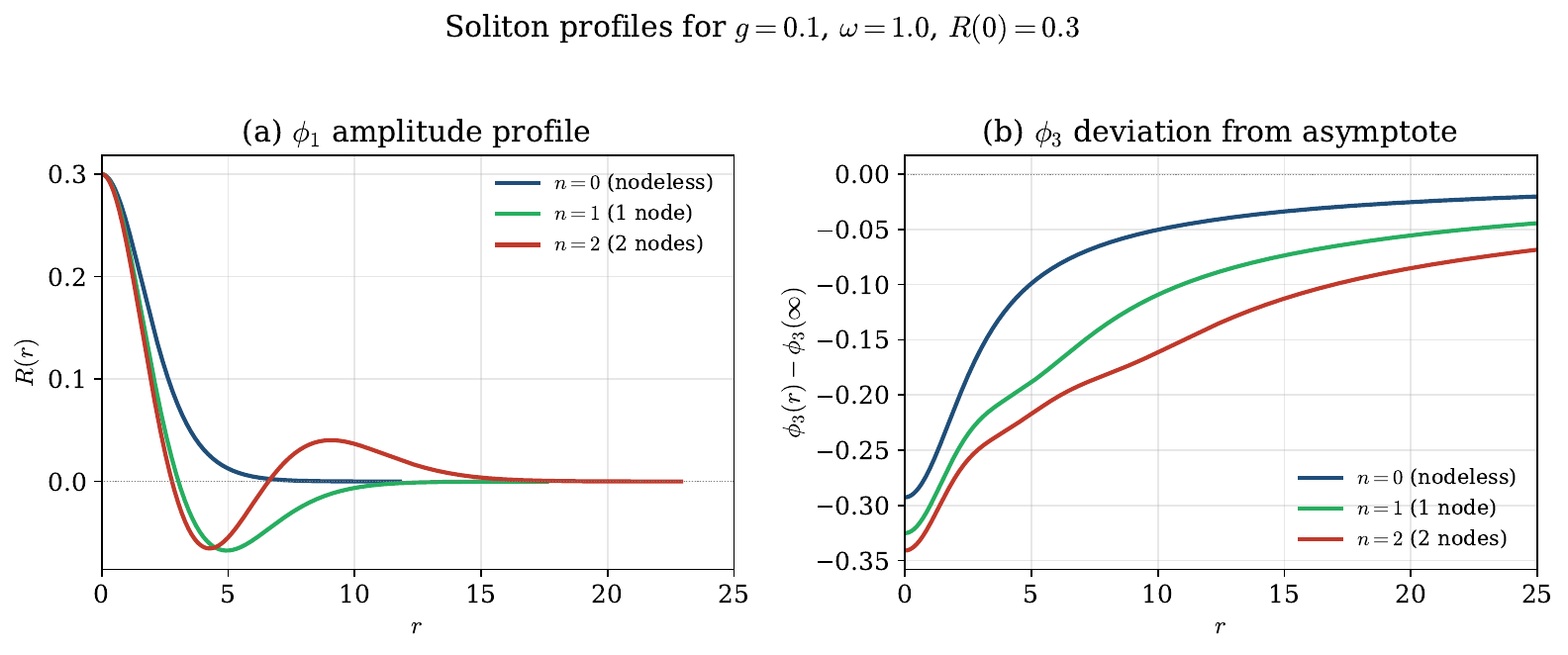}}
    \caption{Numerical solutions for: (a) the radial amplitude $R(r)$
      of the field $\phi_1$; (b) the deviation of the field $\phi_3$
      (equivalently, $S(r)$ in eq.~(\ref{R3shift})) from its
      asymptotic value at large $r$. Parameters are fixed as in
      eq.~(\ref{params}).}
    \label{fig:profiles}
  \end{center}
\end{figure}

Results for the first three states are shown in
fig.~\ref{fig:profiles}, for the parameter choices
\begin{equation}
  y=1,\quad g=0.1,\quad \omega=1,\quad a=0.3.
  \label{params}
\end{equation}
One sees that the amplitude functions $R(r)$ for the $\phi$ field,
corresponding to different critical values $b_*$, indeed constitute
profiles with increasing numbers of nodes. In all cases, the field
$\phi_3$ monotonically increases towards its asymptotic value, which
happens at progessively higher $r$ as the number of nodes
increases. The slight wobbles in the $\phi_3$ curves for more complex
solutions are due to the $\phi_3$ field responding to the influence of
an $R(r)$ field with more oscillations / nodes. In
table~\ref{tab:charges}, we collect the critical values $b_*$ of the
shooting parameter for the solutions of fig.~\ref{fig:profiles}, as
well as their charges and energies. The latter increase with node
number as claimed above.
\begin{table}
  \begin{center}
    \begin{tabular}{c|c|c|c}
      $n$ & $b_*$ & $Q$ & $E$\\
      \hline
      0 & -0.13871385 & 5.8056 & 7.7352\\
      1 & -0.17986543 & 12.7248 & 16.1410\\
      2 & -0.19886936 & 19.6143 & 24.2348
    \end{tabular}
    \caption{Data characterising the soliton solutions of
      fig.~\ref{fig:profiles}, where $n$ is the number of nodes, $b_*$
      the critical value of the shooting parameter (value of $S(r)$ at
      $r=0$), $Q$ the charge, and $E$ the energy.}
    \label{tab:charges}
  \end{center}
\end{table}

\section{Stability of localised solutions}
\label{sec:stability}

\begin{figure}
  \begin{center}
    \scalebox{0.6}{\includegraphics{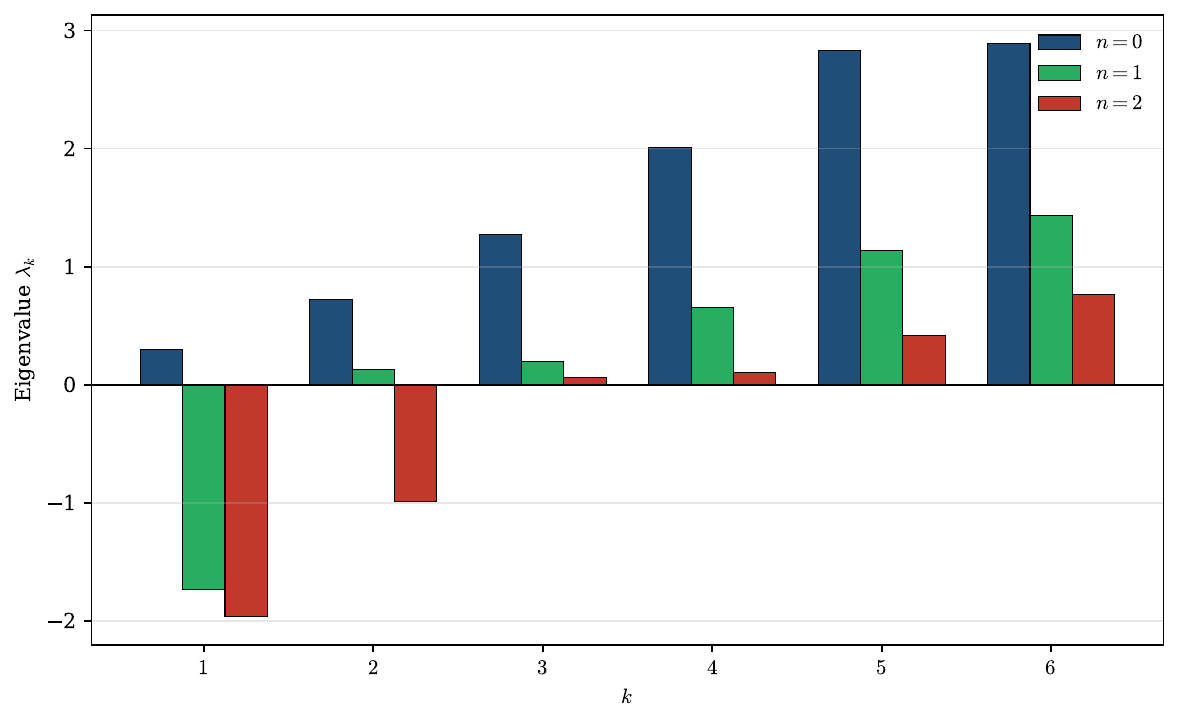}}
    \caption{Eigenvalues of the characteristic equation of
      eq.~(\ref{eigen2}), describing second-order pertubations of
      soliton solutions. Results are shown for the solutions of
      fig.~\ref{fig:profiles}, with node number $n$. Negative
      eigenvalues correspond to unstable modes.}
    \label{fig:eigenvalues}
  \end{center}
\end{figure}

Having verified the existence of localised time-dependent solutions to
biadjoint scalar field theory, let us now examine their stability. To
do this, one starts with the observation that the found soliton
solutions are stationary points of the functional of
eq.~(\ref{Eomega}), which when written in terms of the fields $R(r)$
and $S(r)$ becomes
\begin{equation}
  E_\omega=\int d^3\vec{x}\Big[
    |\nabla R|^2+2|\nabla S|^2+8\beta R^2S+32gR^2S^2+4gR^4
    \Big].
  \label{Eomega2}
\end{equation}
Next, one may perturb around a given solution, and evaluate the
functional
\begin{equation}
  E[R(r)+\rho(r),S(r)+\sigma(r)]=E[R(r),S(r)]+\delta^{(1)} E_\omega
  +\delta^{(2)}E_\omega+\ldots,
  \label{Eperturb}
\end{equation}
where $\delta^{(n)}E_\omega$ captures contributions that are $n^{\rm
  th}$-order in the small deviations $(\rho(r),\sigma(r))$. The
first-order deviation $\delta^{(1)}E_\omega$ vanishes from the
equations of motion, and an explicit calculation reveals the
second-order deviation to be
\begin{equation}
  \delta^{(2)}E_\omega=4\pi\int_0^\infty dr r^2
  \left(\begin{array}{cc}\rho & \sigma\end{array}\right){\cal H}
    \left(\begin{array}{c}\rho \\ \sigma\end{array}\right),    
    \label{delta2}
\end{equation}
where we have defined the matrix operator
\begin{equation}
  {\cal H}=\left(\begin{array}{cc}
        -\nabla^2+8\beta S+32gS^2+24gR^2 & 8\beta R+64gRS\\
        8\beta R+64 g RS & -2\nabla^2+32gR^2
  \end{array}\right)
  \label{Hdef}
\end{equation}
and integrated by parts in the derivative terms. For numerical
purposes, it is convenient to define the vector
\begin{equation}
  \vec{u}=(u_R,u_S)=(r\rho,r\sigma),
  \label{uRuS}
\end{equation}
such that eqs.~(\ref{delta2}, \ref{Hdef}) become
\begin{equation}
  \delta^{(2)}E_\omega=4\pi\int_0^\infty dr 
  \vec{u}^{\rm T} \tilde{{\cal H}}
  \vec{u},\quad \tilde{{\cal H}}=
  \left(\begin{array}{cc}
    -\partial_r^2+8\beta S+32gS^2+24gR^2 & 8\beta R+64gRS\\
        8\beta R+64 g RS & -2\partial_r^2+32gR^2
  \end{array}\right).
    \label{delta2b}
\end{equation}
Stability of a given solution is a statement about the eigenvalues
$\{\lambda_n\}$ of $\tilde{\cal H}$, defined via the differential
equation
\begin{equation}
  \tilde{\cal H}\vec{u}_n=\lambda_n\vec{u}_n,
  \label{eigen}
\end{equation}
where $\{\vec{u}_n\}$ are the corresponding eigenfunctions. If all
eigenvalues are positive, then a given stationary point genuinely
minimises the energy-charge functional, and the solution is stable. If
at least one eigenvalue is negative, then the solution is unstable,
and can decay to a state with the same charge (n.b. the latter is
conserved) but with lower energy. To find the eigenvalues, one may
discretise space into a grid of $r$ values
\begin{equation}
  r_i=i\Delta r,\quad i=1,\ldots, N,
  \label{rivals}
\end{equation}
and define the vector of perturbation values at each point
\begin{equation}
  \vec{x}=(u_R(r_1),\ldots, u_R(r_N),u_S(r_1),\ldots, u_S(r_N)).
  \label{xvec}
\end{equation}
The operator $\tilde{\cal H}$ acting on all field values can then be
represented via the matrix
\begin{equation}
  {\bf H}=\left(\begin{array}{cc}
    -{\bf D}+{\rm diag}(V_{RR}(r_1),\ldots V_{RR}(r_N)) &
    {\rm diag}(V_{RS}(r_1),\ldots V_{RS}(r_N))\\
    {\rm diag}(V_{RS}(r_1),\ldots V_{RS}(r_N)) &
    -2{\bf D}+{\rm diag}(V_{SS}(r_1),\ldots,V_{SS}(r_N))
  \end{array}\right),
  \label{Hmat}
\end{equation}
where
\begin{equation}
  {\bf D}=\frac{1}{(\Delta r)^2}\left(\begin{array}{cccc}
    2 & -1 & 0 & \ldots\\
    -1 & 2 & -1 & \ldots\\
    0 & -1 & 2 & \ldots\\
    \vdots & \vdots & \vdots & \ldots
  \end{array}\right)
  \label{Dmat}
\end{equation}
is the finite difference expression for the second derivative
operator, and
\begin{equation}
  V_{RR}=8\beta S+32gS^2+24gR^2,\quad
  V_{RS}=8\beta R+64gRS,\quad V_{SS}=32gR^2.
  \label{VRR}
\end{equation}
The differential equation of eq.~(\ref{eigen}) now becomes a
conventional matrix characteristic equation
\begin{equation}
  {\bf H}\vec{x}_n=\lambda_n\vec{x}_n,
  \label{eigen2}
\end{equation}
from which one may in principle extract the eigenvalues. However,
there is a further complication in that the general perturbations
$u_R(r)$ and $u_S(r)$ allowed in the above argument are not
necessarily such as to respect the charge conservation constraint
$\delta Q=0$. This generically gives rise to a negative eigenvalue,
corresponding to the fact that unconstrained solutions can lower their
energy by reducing their charge. Upon perturbing a given solution, the
corresponding change in charge is, from eqs.~(\ref{Qdef}, \ref{uRuS}),
\begin{equation}
  \delta Q=16\pi\omega \int dr r R u_R,
  \label{deltaQ}
\end{equation}
which identifies the charge-changing direction in the field space
acted on by $\tilde{\cal H}$ as
\begin{equation}
  \vec{q}\propto \left(\begin{array}{c} r R(r)\\0\end{array}\right).
    \label{qdef}
\end{equation}
One may then project out this direction by replacing the operator
$\tilde{\cal H}$ according to
\begin{equation}
  \tilde{\cal H}\rightarrow {\cal P}\tilde{\cal H}{\cal P},
  \label{Hreplace}
\end{equation}
where
\begin{equation}
  {\cal P}=I-\hat{\vec{q}}\hat{\vec{q}}^{\rm T},\quad
  \hat{q}=\frac{\vec{q}}{|\vec{q}|}
  \label{PIdef}
\end{equation}
is a projection operator that keeps only degrees of freedom orthogonal
to $\vec{q}$. In the discretised eiganvalue problem, this translates
to a matrix operator ${\bf P}$ defined by applying a similar equation
to eq.~(\ref{PIdef}), in terms of a vector
\begin{equation}
  \vec{q}=(r_1 R(r_1),\ldots,r_N(R_N),0,\ldots 0).
  \label{qdef2}
\end{equation}
The resulting eigenvalue problem can then be solved for each distinct
solution for $R(r)$ and $S(r)$. Results for the solutions of
fig.~\ref{fig:profiles} are shown in fig.~\ref{fig:eigenvalues}. For
the lowest energy state, all eigenvalues are positive, implying that
this state is indeed stable under small perturbations. For the higher
states, negative modes appear, such that these states are unstable
with respect to small perturbations. This indicates that such states
can lower their energy by transitioning to the nodeless state with the
same $Q$. In general, the number of negative eigenvalues is the same
as the node number, which is a known feature of Q-ball solutions in
other theories.  

The above arguments address stability under small perturbations. In
fact, in quartic theory with both the cubic and quartic terms turned
on, there is another source of instability due to the fact that the
true vacuum of the theory does not correspond to
$\phi=\phi_3=0$. Rather, the potential of eq.~(\ref{Vdef}) is
minimised at
\begin{equation}
  (|\phi|^2,\phi_3)=\left(\frac{y^2}{16g^2},-\frac{y}{8g}\right).
  \label{minvals}
\end{equation}
Whilst the charge $Q$ protects the soliton solutions found here
against decay to the false vacuum of the theory, they are not
ultimately protected against decay of the false vacuum, as the latter
process proceeds by entirely different non-perturbative processes
(i.e. bubble nucleation of the true vacuum). A special case in which
the solitons are completely stable against vacuum decay is the case
$y=0$ (i.e. complete absence of the cubic interaction). Conversely, a
situation in which the vacuum is even more unstable -- in that the
potential energy is unbounded from below -- is when the quartic
interaction is missing, and the cubic interaction alone is present. We
have explicitly verified that similar non-topological solitons to
those presented in fig.~\ref{fig:profiles} exist even in the purely
cubic case, and that the number of negative eigenvalues for the first
three states at a given $\omega$ follows the same qualitative pattern
as in fig.~\ref{fig:eigenvalues}. Again, this corresponds to the fact
that stability under small perturbations of the solution is not
sensitive to the types of large non-perturbative transition that leads
to decay of the false vacuum.

In the above analysis, we have focused only on perturbations that
preserve the form of the ansatz of eqs.~(\ref{Phiansatz4}, \ref{c123},
\ref{phidef}), as encoded in the reduced Lagrangian of
eq.~(\ref{Ldef}). This is a consistent truncation of the theory, in
that time evolution does not take solutions outside of this particular
subsector. Furthermore, this stability analysis has allowed us to make
contact with existing literature on Q-ball solutions. In principle,
however, one could also check stability against perturbations that go
outside this subsector. Whilst this would be interesting, it is not in
fact necessary for exploring double copy applications of the solutions
presented here. 

\section{Discussion}
\label{sec:discuss}

In this paper, we have continued the programme of work established in
refs.~\cite{White:2016jzc,DeSmet:2017rve,Bahjat-Abbas:2018vgo,Armstrong-Williams:2025spu,Armstrong-Williams:2022apo},
of finding non-linear solutions of biadjoint scalar field theory. The
latter has been widely studied due to its special role in the double
copy relating scattering amplitudes and classical solutions in gauge,
gravity and related theories. No fully non-perturbative definition of
this correspondence exists, and the ultimate scope and remit of the
double copy is unknown. This motivates the study of the
non-perturbative structure of biadjoint theory, which may well provide
clues about how to construct a non-linear double copy, if it
exists. As noted in
refs.~\cite{White:2016jzc,DeSmet:2017rve,Bahjat-Abbas:2018vgo,Armstrong-Williams:2025spu,Armstrong-Williams:2022apo},
nonlinear biadjoint solutions may also be of interest in themselves,
as there may be physical systems (e.g. in condensed matter) that are
themselves described by (generalisations of) biadjoint theory.

The solutions derived in this paper go substantially beyond any
previously constructed objects. They are localised and finite, which
is possible due to their time-dependence. Furthermore, the ansatz used
to construct them is such that they can be straightforwardly embedded
into any choice of non-abelian colour groups. The latter fact is
particularly important for studies of the double copy: colour
structure is removed when proceeding from gauge theory to gravity (as
it must be), which means that non-perturbative double copies, if
possible, must be independent of the choice of gauge group. Our
solutions carry a U(1) charge corresponding to rotations in colour
space about a chosen axis corresponding to an element of the Cartan
subalgebra. This charge protects our solutions against decay to the
trivial solution, and for that reason they are closely related to
non-topological solitons that are known in other scalar theories
e.g. Q-balls. Notably, the solutions are still ultimately unstable in
most of the space of couplings $(y,g)$, due to the unstable nature of
the vacuum. As has been noted for previous
solutions~\cite{White:2016jzc,DeSmet:2017rve,Bahjat-Abbas:2018vgo,Armstrong-Williams:2025spu,Armstrong-Williams:2022apo},
the instability of the vacuum is not necessarily a problem for
potential double copy applications, given that pathological behaviour
can often be remedied when moving between theories. In any case, there
is a regime (that of zero cubic coupling $y$) where the topological
solitons discovered herein do not suffer this fate.

Our results bolster the view that biadjoint scalar theory has a rich
non-perturbative structure, and we hope that they in turn contribute
to a wider discussion of this fascinating theory, and its close
cousins.


\section*{Acknowledgments}

This work has been supported by the UK Science and Technology
Facilities Council (STFC) Consolidated Grant ST/X00063X/1
``Amplitudes, Strings and Duality”. KAW is supported by a
Royal Society Career Development Fellowship.

\bibliography{refs}
\end{document}